\documentclass[aps,prd,reprint,showpacs,superscriptaddress,twocolumn]{revtex4-2}
\usepackage{amsfonts}
\usepackage{amsmath}
\usepackage{amssymb}
\usepackage{dsfont}
\usepackage{hyperref}
\usepackage{graphicx}
\usepackage{float}
\usepackage[usenames,dvipsnames]{xcolor}
\usepackage[normalem]{ulem}
\usepackage{times}
\usepackage{braket}
\usepackage{booktabs}
\usepackage{caption}
\usepackage{subcaption}

\usepackage{natbib}
\usepackage{physics}
\usepackage{mathtools}

\newcommand{\orcid}[1]{\href{https://orcid.org/#1}
	{\includegraphics[width=7pt]{orcid.png}}}

\hypersetup{
	colorlinks=true,linkcolor=blue,citecolor=blue,
	filecolor=blue,urlcolor=blue,breaklinks=true
}
\RequirePackage{color}

\begin{document}
	
	\date{\today}
	
	\title{Thermodynamic properties of an electron gas in a two-dimensional quantum dot: an approach using density of states}
	
	\author{Lu\'{i}s Fernando C. Pereira}
	\email{luisfernandofisica@hotmail.com}
	\affiliation{Departamento de F\'{\i}sica, Universidade Federal do Maranh\~{a}o, 65085-580 S\~{a}o Lu\'{\i}s, Maranh\~{a}o, Brazil}
	
	\author{Edilberto O. Silva}
	\email{edilberto.silva@ufma.br}
	\affiliation{Departamento de F\'{\i}sica, Universidade Federal do Maranh\~{a}o, 65085-580 S\~{a}o Lu\'{\i}s, Maranh\~{a}o, Brazil}
	
	\begin{abstract}
		Potential applications of quantum dots in the nanotechnology industry make these systems an important field of study in various areas of physics. In particular, thermodynamics has a significant role in technological innovations. With this in mind, we studied some thermodynamic properties in quantum dots, such as entropy and heat capacity, as a function of the magnetic field over a wide range of temperatures. The density of states plays an important role in our analyses. At low temperatures, the variation in the magnetic field induces an oscillatory behavior in all thermodynamic properties.
		The depopulation of subbands is the trigger for the appearance of the oscillations. 
	\end{abstract}
	
	\maketitle

\section{Introduction}
\label{intro}

The study of the two-dimensional electron gas (2DEG) is crucial for understanding physical phenomena in semiconductor materials, offering valuable insights into charge transport \cite{JMMM.2007.310.2277}, high-mobility phenomena \cite{PRB.2023.107.195406}, and quantum properties in nanostructures \cite{PRB.2016.93.235411,SSP.1991.44.1}, all pivotal for advancements in electronics and nanotechnological devices \cite{4601361,NN.2018.13.915,N.2016.27.365701}. 2DEG has been used to describe the physics of various mesoscopic models to study the properties of quantum dots \cite{PRB.2022.105.155302,PRB.2023.108.075303,PLA.2016.380.3847,NC.2023.14.4876}, quantum rings \cite{AoP.2023.459.169547,AdP.2023.535.202200371,FBS.2022.63.64,FBS.2022.63.58,PE.2021.132.114760}, and other effective theories \cite{AdP.2018.530.1800112,PLA.2015.379.2110,PLA.2015.379.907}. Over the years, extensive research has focused on the thermodynamics of a two-dimensional electron gas and its dependence on magnetic fields. Experimental data reveal that equilibrium properties at low temperatures exhibit oscillations depending on the magnetic field strength \cite{PRL.1985.54.1820,PRB.1992.45.4384,PRB.2002.65.245315}. Theoretically, when a magnetic field is applied perpendicular to the 2DEG plane, degenerate energy levels, the Landau levels, are formed \cite{BOOK.Heinzel.2008}. This characteristic is the basis for the oscillatory behavior of all thermodynamic properties, such as the oscillations observed in heat capacity, chemical potential, and de Haas-van Alphen oscillations (dHvA effect) in magnetization \cite{PRL.1983.51.1700,SSC.1984.50.537,SSC.2008.146.487,SM.2013.59.60,EPJP.2023.138.983}.
Chemical potential plays an important role in understanding oscillations. Hence, importance is given to the study of the dependence of chemical potential on the magnetic field and temperature \cite{Vagner.2006.3.102}. From a practical point of view, the contribution of studies on the thermodynamics of mesoscopic systems is vital for technological development. For example, heat management is an important factor in designing elements used in electronic circuits \cite{HTE.2007.28.255,SMO.2022.65.297}. In this sense, we must know how a device behaves with temperature variations and external fields. Studies on thermodynamics also play a vital role in environmental issues.
Nowadays, environmental concerns stimulate scientific research in search of proposals for efficient management of natural resources. In this context, the study of the 
dependence of temperature on the magnetic field becomes very important since it is possible to replace traditional refrigeration technology with a cleaner alternative  The dependence of temperature on the magnetic field is known as the magnetocaloric effect.

In recent decades, semiconductor quantum dots have emerged as pivotal entities in condensed matter physics and quantum technology \cite{quantumdots2023}. These nanoscale structures, confined in all three spatial dimensions, exhibit remarkable properties that have garnered substantial attention due to their versatile applications in various fields of physics and technology. Semiconductor materials like GaAs have been extensively explored for quantum dot fabrication \cite{S.1998.32.343,AFM.2011.21.869,JJAP.2000.39.L79}, owing to their tunable electronic and optical properties \cite{bhattacharya2017semiconductor}. Information transport in semiconductor quantum dots forms a cornerstone in quantum computing and communication. The discrete energy levels and controllable confinement potential of quantum dots make them promising candidates for qubits \cite{CPB.2018.27.020305}, the fundamental units of quantum information \cite{PRA.1998.57.120}.

The investigation into the thermodynamic properties of semiconductor quantum dots, as pursued in this study, plays an important role in comprehending the behavior of these nanostructures under diverse conditions \cite{PRB.1992.46.12773,JKPS.1995.28.132,PRB.2008.78.045321,JAP.2012.112.083514,PE.2018.103.464,Entropy.2018.20.557,Entropy.2019.21.512,AdP.2019.531.1900254}. Exact energy spectra and wave functions are obtained analytically when a parabolic potential models a quantum dot. In contrast to the Landau levels, accidental degenerations occur when the magnetic field varies \cite{RMP.2002.741283}. This behavior is responsible for much of the new physics uncovered in the quantum dot systems \cite{SSR.2001.41.1}.
Determining electronic states, internal energy, magnetization, Helmholtz free energy, specific heat, and entropy contributes to the fundamental understanding required for harnessing these materials in various technological applications. 
A comprehensive understanding of their physical properties is imperative for exploiting their potential across a spectrum of quantum-enabled technologies.

The structure of our article is presented as follows. In Section \ref{sec:model}, we present the model that describes the motion of an electron with effective mass $\mu$ and electric charge within a quantum dot in the presence of a uniform magnetic field along the z-direction. We derive the Schr\"{o}dinger equation for the system and determine the eigenvalues of energy and corresponding wave functions. Section \ref{sec:cp} is dedicated to studying the chemical potential of the system, particularly considering the scenario of a semiconductor quantum dot with multiple electrons. Initially, we explore the Fermi energy and its relationship with the physical parameters involved. From these analyses, we derive the density of states, a very important parameter that allows us to expand our field of study. Section \ref{sec:Mag}, we examine the magnetization as a function of the magnetic field at different temperatures. Section \ref{sec;entropy} is dedicated to the study of the system's entropy. We take the opportunity to present an important physical phenomenon, the magnetocaloric effect, which consists of the variation in temperature as a function of the magnetic field. In Section \ref{sec:hc}, we study the heat capacity of a two-dimensional electron gas in a quantum dot.
Our conclusions are given in Section \ref{sec:Conc}.  
     
\section{Description of the model}
\label{sec:model}

In this section, we describe the model of an electron with effective mass $\mu$ and charge $e$ confined  by a radial potential  $V\left(r\right)$ and
under the influence of a uniform magnetic field \( \mathbf{B}\) along the z-direction. The motion of the particle is governed by the Schr\"{o}dinger equation with minimal coupling
\begin{equation}
\left[\frac{1}{2\mu }\left( \mathbf{p}-e\mathbf{A}\right) ^{2}+V\left(
r\right)\right]\psi\left(r,\varphi\right)=E\psi\left(r,\varphi\right),  
\label{Sch}
\end{equation}
where $\mathbf{p}$ is the momentum operator and $\mathbf{A}$ denotes the vector potential. In this formulation, we adopt the symmetric gauge vector potential given by
\begin{equation}
\mathbf{A}=\frac{Br}{2}\hat{\boldsymbol{\varphi}},
\label{Eq:calibre}
\end{equation}
where \(B \) represents the magnetic field strength. Additionally, the confinement potential is given by \cite{SST.1996.11.1635} 
\begin{equation}
V\left( r\right)=\frac{a_{1}}{r^{2}}+a_{2}r^{2}-2\sqrt{a_{1}a_{2}}.
\label{PotRad}
\end{equation}
This potential has a minimum at $r_{0}=\left(a_{1}/a_{2}\right)^{1/4}$. Furthermore, it can be shown that $\omega_{0}=\sqrt{8a_{2}/\mu}$, which defines the strength of the transverse confinement. It is known that any variation in parameters $a_{1}$ and $a_{2}$ implies changes in the spatial distribution of the electron. The radial potential serves as a model for the theoretical definition of a ring with finite width. However, it can also describe other systems, such as a quantum dot with the condition $a_{1}=0$. The radial potential creates an environment where the electron's quantum properties are highlighted, giving rise to phenomena such as confinement and discrete energy level structures. 
Through the model of the interaction of an electron with the magnetic field at a radial potential, we can explore phenomena such as bound-state formation, discrete energy spectra, and spatial confinement effects. These aspects are fundamental for understanding the physics of nanostructured systems and their potential applications in quantum devices.

For the case of an electron confined in a 
quantum dot, the energy eigenvalues and wavefunctions of Eq. (\ref{Sch}) are given, respectively, by
\begin{equation}
E_{n,m}=\left( n+\frac{1}{2}+ \frac{\left|m\right|}{2}\right) \hbar\omega  -
\frac{m}{2}\hbar \omega_{c},  \label{Eq:Enm}
\end{equation}
and
\begin{align}
\psi_{n,m} (r,\varphi ) &= \frac{1}{\lambda }\sqrt{\frac{\left( n+\left\vert
m\right\vert \right) !}{2\pi \,n!\left( \left\vert m\right\vert !\right) ^{2}}}
e^{im\varphi }e^{ -\frac{r^{2}}{4\lambda ^{2}}}\left( \frac{r^{2}}{2\lambda ^{2}}\right) ^{\frac{\left\vert
m\right\vert }{2}}
\notag \\
& \times{\mathrm{M}
\left( -n,\,1+\left\vert m\right\vert, \frac{r^{2}}{2\lambda ^{2}}\right)},
\label{Eq:funcaodeonda}
\end{align}
where $n=0,1,2,\ldots$ is the radial quantum number, which denotes a subband, $m=0,\pm 1,\pm2,\ldots$ is the magnetic quantum number. Also, $M\left(a,c,x\right)$ is the confluent hypergeometric function of the first kind, $\omega = \sqrt{\omega_{0}^{2}+\omega_{c}^{2}}$
is the effective cyclotron frequency, $\omega_{c}=eB/\mu$ is the cyclotron frequency, and $\lambda=\sqrt{\hbar/\left(\mu \omega\right)}$ is the effective magnetic length. 

By using Eq. (\ref{Eq:Enm}), we can show that the minimum energy of a subband is given by
\begin{equation}
E_{n,0}=\left( n+\frac{1}{2}\right) \hbar\omega,
\label{Eq:En0}
\end{equation}
and the separation energy between the bottoms of neighboring subbands is $\hbar \omega$.

In subsequent sections, we study the dependence of some thermodynamic properties on temperature and the magnetic field. For numerical analysis, we consider a sample made of GaAs with the effective mass of the electron $\mu=0.067 \mu_{e}$, where $\mu_{e}$ is the electron mass. We also consider that there are $N=1400$ spinless electrons.
The confinement energy parameter used for this purpose is $\hbar \omega_{0}=0.459$ meV \cite{PRB.1999.60.5626}.  

\section{Chemical potential}
\label{sec:cp}

The Fermi energy corresponds to the energy of the topmost filled level in the $N$ electron system \cite{kittel2005introduction}. Furthermore, a subband at the Fermi energy is only partially occupied. With that in mind, we can define the Fermi energy as
\begin{equation}
E_{F}=\left( n_{max }+\frac{1}{2}\right) \hbar \omega +\delta \hbar \omega
\label{Eq:Fermi 1}
\end{equation}
In this equation, $n_{max}$ corresponds to the highest occupied subband at the Fermi energy, and $0 \leq \delta < 1$. Starting from the idea that there are $N$ electrons distributed in $n_{max}+1$ subbands, we can show that $n_{max}$ and $\delta$ are given, respectively, by
\begin{equation}
n_{max}\equiv [x],\; x=\sqrt{\frac{1}{4}+2\nu}-\frac{1}{2},
\label{Eq:nmax}
\end{equation}
and 
\begin{equation}
\delta =\frac{\nu}{n_{max }+1}-\frac{n_{max}}{2},  
\label{Eq:delta}
\end{equation}
where $[x]$ denotes the largest integer less than or equal to $x$, $\nu=N/\Delta N$, and
\begin{equation}
\Delta N=\left( \frac{2\omega }{\omega _{0}}\right)^{2}, 
\label{Eq:dNn}
\end{equation} 
is the number of states in a subband in energy interval $E_{n,0} \leq E < E_{n+1,0}$. From these results, we can deduce that the density of states of a subband is
\begin{equation}
D\left(E\right)=\frac{\Delta N}{\hbar \omega}.
\label{Eq:DOS}
\end{equation}
The density of states is of fundamental importance in studies of a 2DEG. This parameter is directly derivable
from measurable properties such as magnetization, capacitance, and heat capacity \cite{PRB.1992.45.4384,EPJP.2023.138.983}.

By using the Eqs. (\ref{Eq:Fermi 1})-(\ref{Eq:dNn}), we can also express the Fermi energy as a function of the parameter $\delta$ and the filling factor $\nu$ as 
\begin{equation}
E_{F}=\hbar \omega \sqrt{\left( \frac{1}{2}-\delta \right) ^{2}+2\nu}.  
\label{Eq:Fermi 3}
\end{equation}
It is worth noting that an expression for Fermi energy in a quantum dot as a function of the square root of number electrons $N$ and the frequencies $\omega$ and $\omega_{c}$ was obtained in Ref. \cite{PE.2023.147.115617}. Such a result corresponds to the case when $\delta=0$ in Eq. (\ref{Eq:Fermi 3}). Comparing Eqs. (\ref{Eq:En0}) and (\ref{Eq:Fermi 1}), this particular case tells us that the Fermi energy is located at the bottom of the subband, i.e., when the subband is nearly depleted. These are the points at which the model from Ref. \cite{PE.2023.147.115617} coincides with the exact results.

So far, we have considered the ideal case of zero temperature. However, from the density of states, given by Eq. (\ref{Eq:DOS}), we can obtain the chemical potential and magnetization at any temperature. In addition, other thermodynamic properties that depend on temperature can be accessed, such as the heat capacity of electrons and the system's entropy. In that case, the number of electrons is computed as \cite{kittel2005introduction}
\begin{equation}
N=\displaystyle \sum_{n=0}^{\infty}\int_{E_{n,0}}^{\infty} f\left(E,\chi \right) D\left(E\right)dE.
\label{Eq:N}
\end{equation}
In this equation, $D\left(E\right)$ is given by Eq. (\ref{Eq:DOS}), and $f\left(E,\chi\right)$ is the Fermi-Dirac distribution
function, given by
\begin{equation}
f\left(E,\chi\right)=\frac{1}{1+\exp\left(\frac{E-\chi}{k_{B}T}\right)},
\label{Eq:distribuicao}
\end{equation}
where $\chi$ is chemical potential and $k_{B}$ is the Boltzmann constant. In a system at temperature $T$ containing $N$ electrons, the chemical potential is calculated from Eq. (\ref{Eq:N}). In the limit $T=0$, the chemical potential corresponds to the Fermi energy given by  Eq. (\ref{Eq:Fermi 1}).

\begin{figure}[!h!]
\centering	
\includegraphics[scale=0.55]{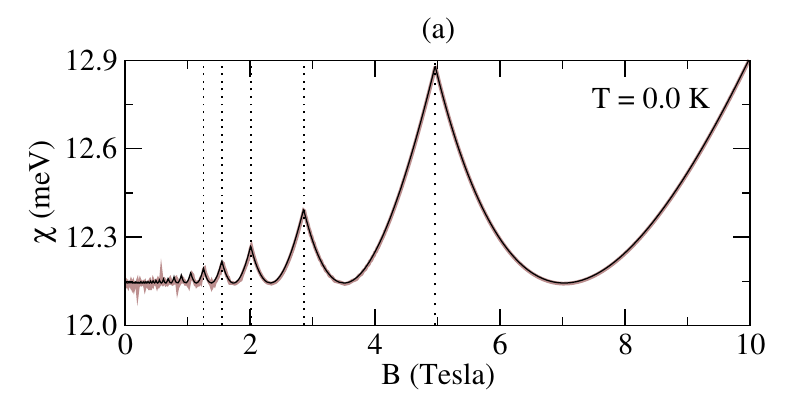}
\includegraphics[scale=0.55]{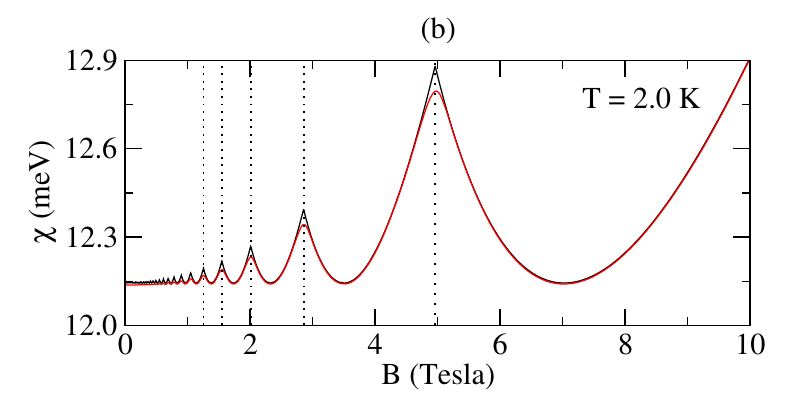}
\includegraphics[scale=0.55]{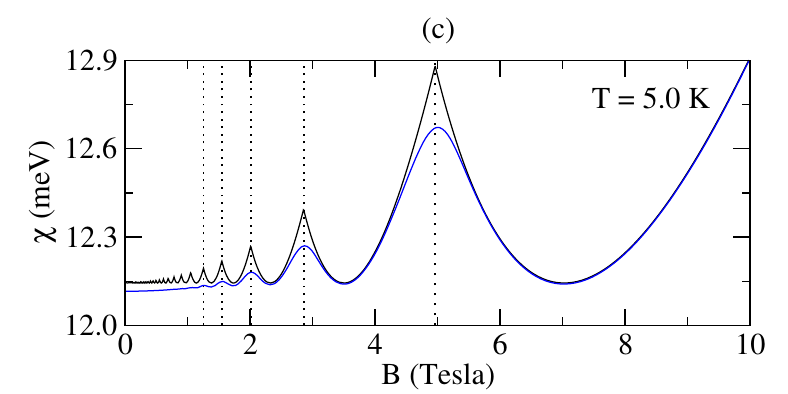}
\caption{(a) The black line corresponds to the chemical potential at $T=0$ computed from Eq. (\ref{Eq:Fermi 1}). To compare with exact results, we also plot the Fermi energy computes self-consistently (line brown). In (b) and (c), the red and blue lines correspond to the chemical potential at finite temperatures obtained from Eq. (\ref{Eq:N}). We plot the chemical potential at $T=0$ (black line) to better visualize the temperature effect.
The dashed lines in (a), (b), and (c) show the position at which subbands with $n=1,2,3,4,5$ are depopulated. The corresponding magnetic fields are obtained from Eq. (\ref{Eq:B-maximos}).}
\label{Fig:fermi}
\end{figure}

Figure \ref{Fig:fermi} shows the dependence of the chemical potential on the magnetic field at different temperatures. In Fig. \ref{Fig:fermi}(a), the black and brown lines correspond to the Fermi energy computed from Eq. (\ref{Eq:Fermi 1}) and self-consistently \cite{AIPA.2014.4.127151}, respectively. As we can see, our approach leads to an excellent agreement on the exact result. Nevertheless, our model considers a continuous density of states, meaning some results are not observed. In Section \ref{sec:Mag}, we will return to this question. The oscillations, consisting of a series of almost parabolas with cusps pointing upward, are attributed to the depopulation of energy subbands when the magnetic field increases. Peaks occur when the Fermi energy is at the bottom of a subband, which, as noted above, occurs with the condition $\delta=0$. Thus, the Eq. (\ref{Eq:delta})
provides
\begin{equation}
B_{peaks}=\frac{\mu\omega_{0}}{e}\sqrt{\frac{N}{2n\left(n+1\right)}-1},
\label{Eq:B-maximos}
\end{equation}
which corresponds to the magnetic field values for the oscillations to take maximum value. Note that $n \neq 0$.
This condition is expected since the Fermi energy does not reach the bottom of the subband with $n=0$.  
Another result that we can obtain from Eq. (\ref{Eq:B-maximos}) is that increasing the confinement intensity $\hbar \omega_{0}$ leads to the shift of the peaks to higher magnetic fields. This also increases the Fermi energy. However, there is no increase in the number of oscillations. The shift of peaks to higher magnetic fields also occurs if the number of electrons $N$ increases. In this case, there is also an increase in the number of oscillations. Equation (\ref{Eq:dNn}) shows that $\Delta N$  increases with the square of the magnetic field fields. 
The $\Delta N$ parameter is small for weak magnetic fields, so many occupied subbands are required to accommodate 1400 electrons. Furthermore, the depopulation of subbands is rapid. For strong magnetic fields, it is the opposite; namely, $\Delta N$ is large such that few occupied subbands are needed, and depopulation occurs over a larger magnetic field range. This explains both the increase in the amplitude and the increased period of the oscillations. Exactly when the subband with $n=1$ is completely depopulated, $\Delta N=1400$ such that all electrons can be fitted into the subband with $n=0$. In this case, there are no more oscillations. The quantum dot states tend to the usual Landau quantization in the range of strong magnetic fields. 
For finite temperatures, the Figs. \ref{Fig:fermi}(b) and \ref{Fig:fermi}(c) show that the oscillations are softened and washed out. Furthermore, we can observe the decrease in chemical potential at finite temperatures. Indeed, as the number of electrons $N$ must remain constant, the chemical potential must decrease with increasing temperature. In the high-temperature regime, the role played by subband depopulation is less important.

\section{Magnetization}
\label{sec:Mag}

At zero temperature, the magnetization of a system containing
a fixed number $N$ of the electron is given by $\mathcal{M}=-\partial U/\partial B$, where $U$ is the total internal energy. We start from the idea that each subband contributes to an energy given by
\begin{equation}
U_{n}=\int_{E_{n,0}}^{E_{F}} E D\left(E\right)dE.
\label{Eq:UnRelacao}
\end{equation}
Since the energy density is independent of the energy range considered, the total internal energy is easily computed: 
\begin{equation}
U=\frac{1}{2}D\left(E\right)\displaystyle \sum_{n=0}^{n_{max}}\left(E_{F}^{2}-E_{n,0}^{2}\right).
\label{Eq:Energia}
\end{equation}
\begin{figure}[!t!]
\centering	
\includegraphics[scale=0.55]{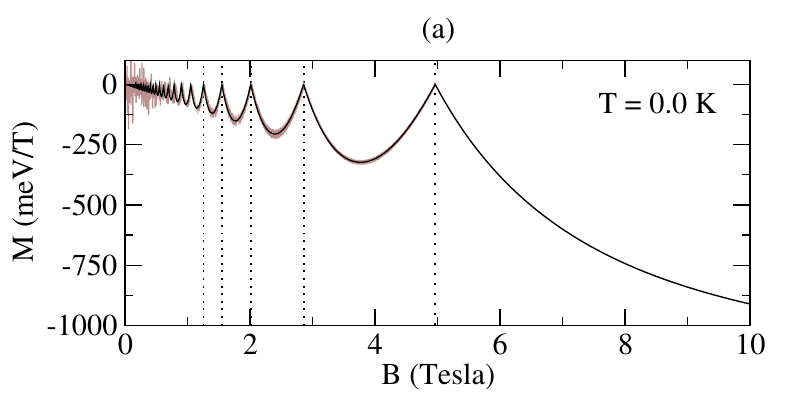}
\includegraphics[scale=0.55]{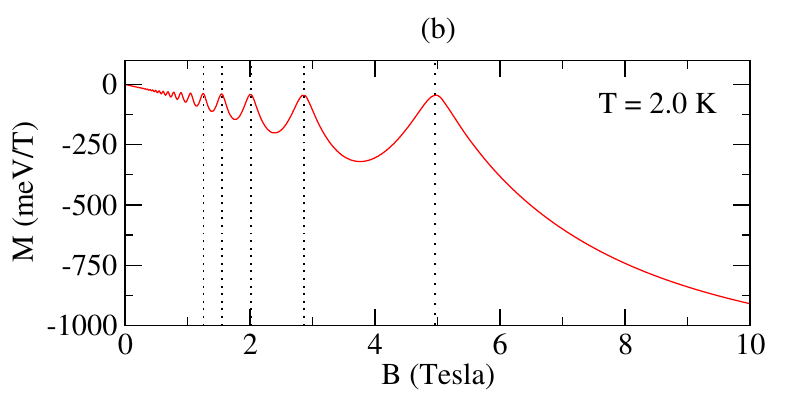}
\includegraphics[scale=0.55]{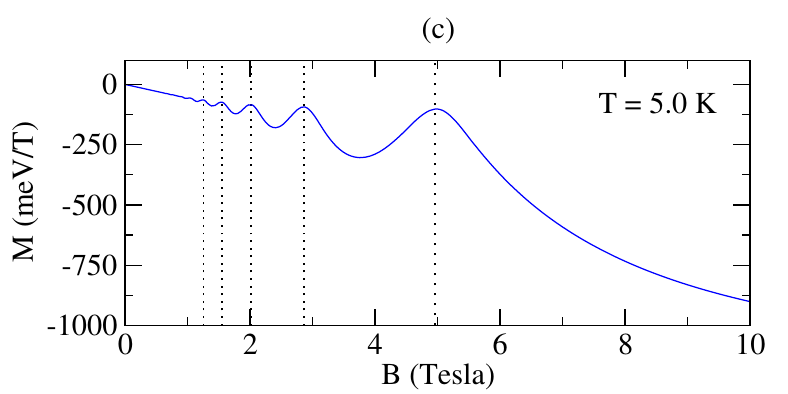}
\caption{(a) The black line corresponds to the magnetization at $T=0$ computed from Eq. (\ref{Eq:Mag1}). We also plot the exact magnetization results (line brown) for comparison. In (b) and (c), the red and blue lines correspond to the magnetization at finite temperatures obtained from Eq. (\ref{Eq:Mag3}). 
The dashed lines in (a), (b), and (c) show the position at which subbands with $n=1,2,3,4,5$ are depopulated. The corresponding magnetic fields are obtained from Eq. (\ref{Eq:B-maximos}).}
\label{Fig:magnetizacao}
\end{figure}
As $\mathcal{M}=-\partial U/\partial B$, we can show that the magnetization is given by
\begin{equation}
\mathcal{M}=-\mathcal{M}_{0}\left(\frac{\omega _{c}}{\omega }\right)\left(\frac{6U-4NE_{F}}{\hbar \omega}\right),
\label{Eq:Mag1}
\end{equation}
where $\mathcal{M}_{0}=\hbar e/2\mu$. 

For magnetic fields in which only the subband with $n=0$ is occupied, we can show from Eq. (\ref{Eq:Mag1}) that the magnetization is given by
\begin{equation}
\mathcal{M}=-N\mathcal{M}_{0}\left( \frac{\omega _{c}}{\omega }\right)
\left( 1-\delta \right).
\label{Eq:Mag2}
\end{equation}
For strong magnetic fields, $\omega \rightarrow \omega_{c}$, while $\delta \rightarrow 0$. Consequently, in the limit of strong magnetic fields, we verify that the absolute value of magnetization tends to the value $N\mathcal{M}_{0}$.

The magnetization for non-zero temperatures is given by 
\begin{equation}
\mathcal{M}=-\frac{\partial F}{\partial B},
\label{Eq:Mag3}
\end{equation}
where $F$, the free energy, is computed as
\begin{equation}
F=N\chi-k_{B}T\displaystyle \sum_{n=0}^{\infty}\int_{E_{n,0}}^{\infty} \ln\left[1+\exp\left(\frac{\chi-E}{k_{B}T}\right)\right] D\left(E\right)dE.
\label{Eq:EnergiaLivre}
\end{equation}

Figure \ref{Fig:magnetizacao} shows the dependence of magnetization on the magnetic field at different temperatures. The oscillations correspond to the dHvA effect. In Fig. \ref{Fig:magnetizacao}(a), the black line corresponds to the magnetization computed from Eq. (\ref{Eq:Mag1}). The brown line corresponds to the exact magnetization results. Here, we understand the question raised in Section \ref{sec:cp} about the continuous density of states. It is a well-known result in the literature that in addition to dHvA-type oscillations, there is also a second oscillation in the magnetization of low-dimensional systems. These oscillations are associated with crossing states and are defined as type-AB oscillations \cite{PRB.1999.60.5626}. For low magnetic fields, where $\omega_{c} \ll \omega_{0}$, the AB-type oscillations superimpose on the dHvA-type oscillations. For strong magnetic fields, where $\omega_{c} > \omega_{0}$, is the opposite, AB-type oscillations are suppressed by increasing the magnetic field, while the amplitude of dHvA-type oscillations increases. This is precisely the situation presented by the brown line in Fig. \ref{Fig:magnetizacao}(a). In our model, the density of states is a continuous parameter. As a result, AB-type oscillations are not observed. We emphasize the small amplitudes of these oscillations, particularly in the regime of strong magnetic fields. As a consequence, the amplitude of these oscillations decreases at temperatures lower than those shown in Figs. \ref{Fig:magnetizacao}(b) and \ref{Fig:magnetizacao}(c) \cite{PE.2021.132.114760}.

\section{Entropy}
\label{sec;entropy}

\begin{figure}[t!]
\centering
\includegraphics[scale=0.55]{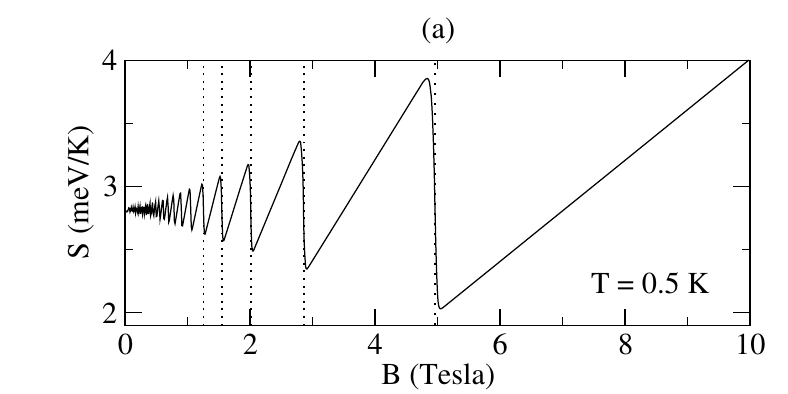}
\includegraphics[scale=0.55]{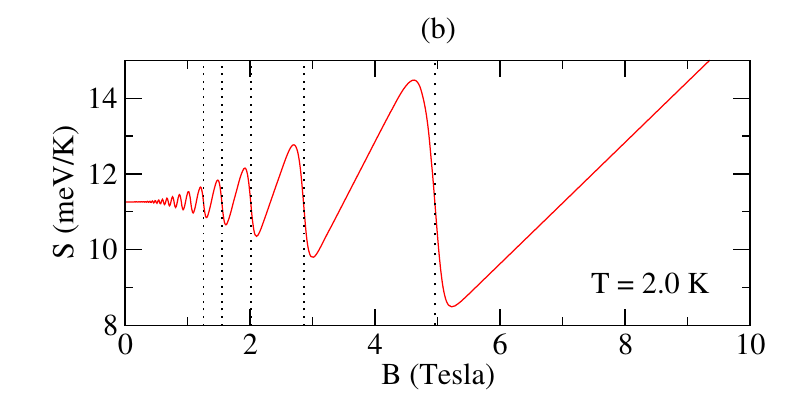}
\includegraphics[scale=0.55]{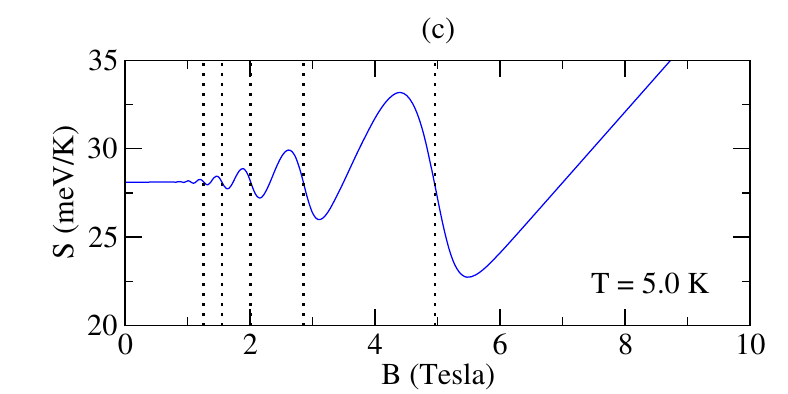}
\includegraphics[scale=0.55]{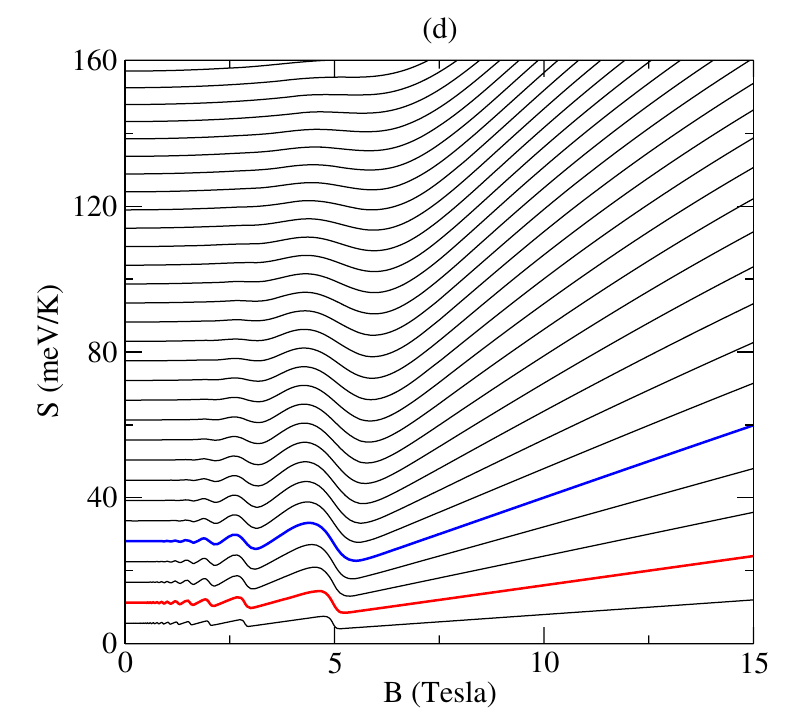}
\caption{Entropy (Eq. (\ref{Eq:entropia})) as a function of the magnetic field in a wide range temperature. The dashed lines in (a) and (b) show the position at which subbands with $n=1,2,3,4,5$ are depopulated. The corresponding magnetic fields are obtained from Eq. (\ref{Eq:B-maximos}). In (c), temperatures vary from $1.0$ K to $30.0$ K, with a range of $1.0$ K. The curves highlighted in red and blue correspond to temperatures of $1.0$ K and $5.0$ K, respectively.}
\label{Fig:Entropia}
\end{figure}

The entropy of the system is computed from $S=-\partial F/\partial T$. Using $F$ given by Eq. (\ref{Eq:EnergiaLivre}), we write the entropy as
\begin{equation}
S=k_{B}\displaystyle \sum_{n=0}^{\infty}\int_{E_{n,0}}^{\infty}
\left[ \ln \left( 1+e^{\frac{\chi -E}{k_{B}T}}\right) +\frac{\frac{E-\chi }{k_{B}T}}{e^{\frac{E-\chi}{k_{B}T}}+1}\right]D\left( E\right) dE.  
\label{Eq:entropia}
\end{equation}

Figure \ref{Fig:Entropia} shows entropy as a function of the magnetic field in different temperatures. 
From Fig. \ref{Fig:Entropia}(a), we can see that the entropy oscillations are almost perfectly sawtoothlike for lower temperatures. The vertical dashed lines, which correspond to the $B$ values given by Eq. (\ref{Eq:B-maximos}), show that at each dHvA period, the entropy is pinned to its value at zero magnetic fields. 
Entropy is assigned to the states of a subband within an energy range $k_{B}T$ of the chemical potential that can be thermally excited to higher levels.
Far from the bottom of a subband, in the ranges of magnetic fields where the chemical potential is equivalent to the Fermi energy, 
the entropy is approximately given by \cite{Book.Yoshioka.2007.147}
\begin{equation}
S \simeq \frac{1}{3}\left( n_{max}+1\right) \pi ^{2}D\left( E\right)
k_{B}^{2}T.
\label{Eq:entropia2}
\end{equation}
As $T\rightarrow 0$, $S\rightarrow 0$, in accordance with the third law of thermodynamics.
On the other hand, near the bottom of a subband, the contribution of thermally excited states of the subband that is being depopulated decreases, which implies a reduction in the system's entropy. When the contribution from the lower subbands prevails, the entropy increases again with the increase in the magnetic field. Although entropy is an increasing function of temperature, the dependence is not always linear. For example, in strong magnetic fields, the rate of entropy change is high at low temperatures; at high temperatures, it is the opposite. This result can be seen in Fig. \ref{Fig:Entropia}(c).

\begin{figure}[!t]
\centering
\includegraphics[scale=0.55]{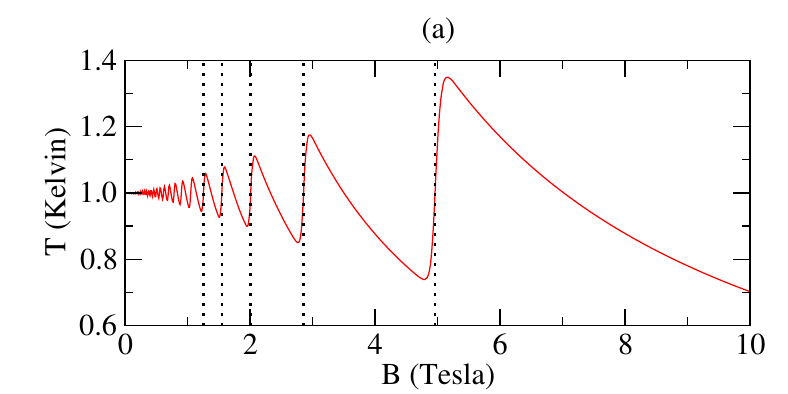}
\includegraphics[scale=0.55]{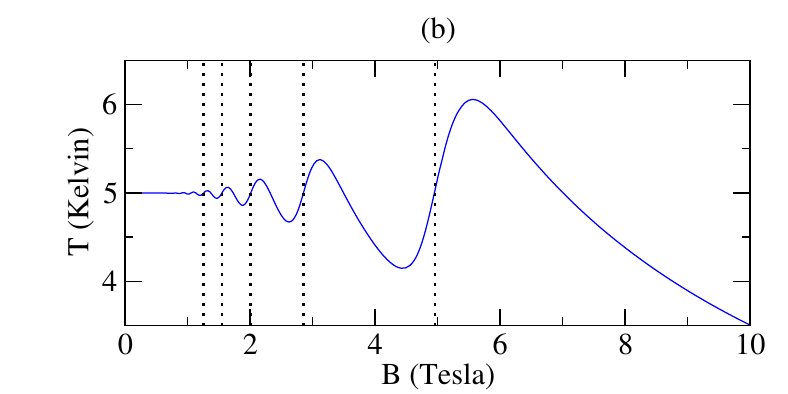}
\caption{Temperature as a function of the magnetic field. The dashed lines in (a) and (b) show the position at which subbands with $n=1,2,3,4,5$ are depopulated. The corresponding magnetic fields are obtained from Eq. (\ref{Eq:B-maximos}).}
\label{Fig:EMC}
\end{figure}

In a thermally isolated system, there is no exchange of energy, and any process that occurs must be carried out adiabatically, i.e., the entropy must remain constant. 
As described above, entropy presents a complex pattern as a function of the magnetic field in different temperature regimes. In this context, for entropy to remain constant for all magnetic field values, the temperature must be a function of the magnetic field; this is the definition of the magnetocaloric effect \cite{SSC.1984.50.537,Entropy.2018.20.557}. 
The magnetocaloric effect is
said to be normal when the temperature increases with the magnetic field. Figure \ref{Fig:EMC} shows that this scenario occurs when the
chemical potential is close to the bottom of a subband. On the other hand, when the temperature decreases with an increase in the magnetic field, we have the inverse magnetocaloric effect. Figure \ref{Fig:EMC} shows that this scenario occurs when the chemical potential is far from the bottom of a subband. In particular, the inverse magnetocaloric effect occurs in the range of strong magnetic fields in which a single subband is occupied. This is also the case when the initial temperature is very high, as we can infer from Fig. \ref{Fig:Entropia}(c). 
As noted by Reis \cite{APL.2011.99.052511}, the above results open doors for applications at quite low temperatures and can be further developed to be incorporated into adiabatic demagnetization refrigerators, as well as sensitive magnetic field sensors.

\section{Heat capacity}
\label{sec:hc}

The heat capacity of the electron gas is given in general as
\begin{equation}
C_{e}=\displaystyle \sum_{n=0}^{\infty}\int_{E_{n,0}}^{\infty} \frac{df\left(E,\chi \right)}{dT} D\left(E\right)\left(E-\chi\right)dE.
\label{Eq:calorespecifico}
\end{equation}
Following the same procedures as in Ref. \cite{SSC.1984.50.537}, we can show that the heat capacity can be written as 
\begin{equation}
C_{e}=k_{B}D\left(E\right)\left(L_{2}-\frac{L_{1}^{2}}{L_{0}}\right),
\label{Eq:calorespecifico2}
\end{equation}
where we define the parameter $L_{r}$ as
\begin{equation}
L_{r}=\displaystyle \sum_{n=0}^{\infty}\int_{E_{n,0}}^{\infty}\frac{\exp\left(\frac{E-\chi}{k_{B}T}\right)}{\left[1+\exp\left(\frac{E-\chi}{k_{B}T}\right)\right]^{2}}\left(\frac{E-\chi}{k_{B}T}\right)^{r}dE.
\label{Eq:L}
\end{equation}

\begin{figure}[!h]
	\centering
	\includegraphics[scale=0.55]{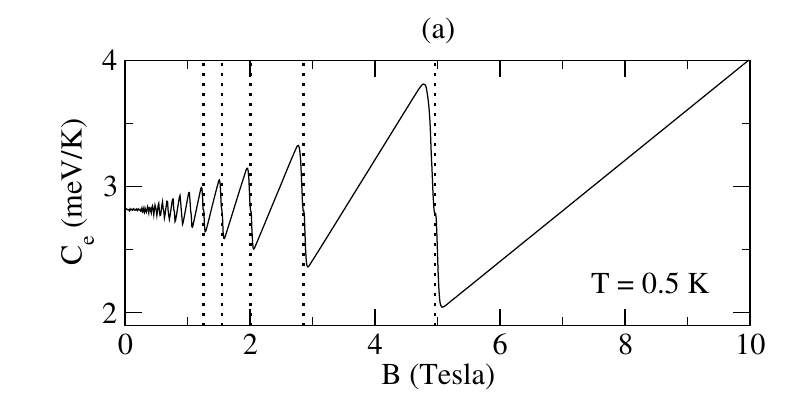}
	\includegraphics[scale=0.55]{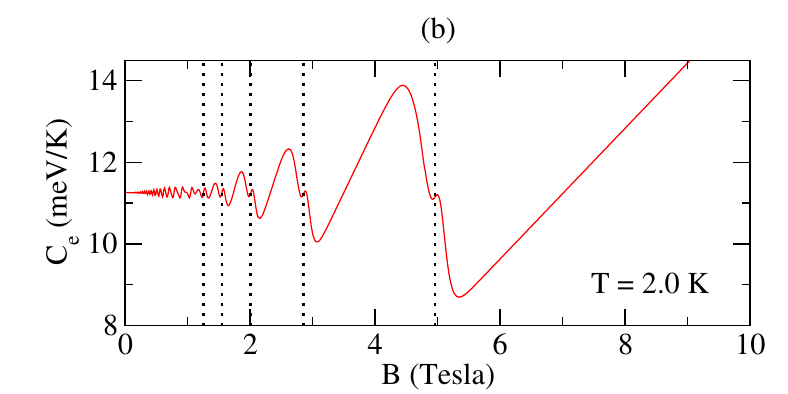}
	\includegraphics[scale=0.55]{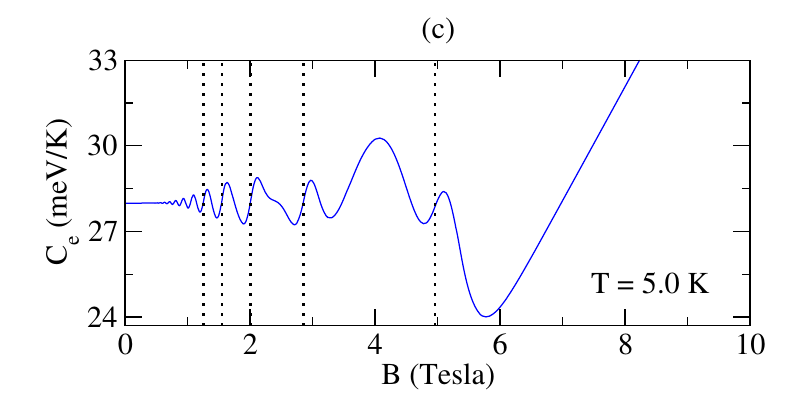}
	\includegraphics[scale=0.55]{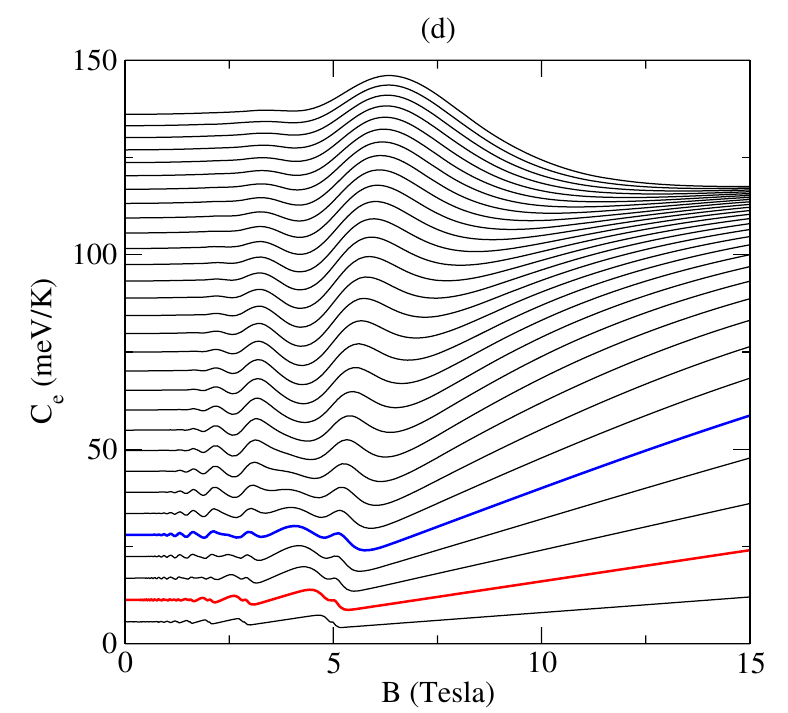}
	\caption{Heat capacity (Eq. (\ref{Eq:calorespecifico2})) as a function of the magnetic field in a wide range temperature. The dashed lines in (a), (b), and (c) show the position at which subbands with $n=1,2,3,4,5$ are depopulated. The corresponding magnetic fields are obtained from Eq. (\ref{Eq:B-maximos}). In (d), temperatures vary from $1.0$ K to $30.0$ K, with a range of $1.0$ K. The curves highlighted in red and blue correspond to temperatures of $2.0$ K and $5.0$ K, respectively.}
	\label{Fig:CalorEspecifico}
\end{figure}

Figure \ref{Fig:CalorEspecifico} exhibits the effect of the interplay between temperature and magnetic fields on the heat capacity. Figure \ref{Fig:CalorEspecifico}(a) shows that, in the low-temperature regime, heat capacity has a similar pattern to entropy, as we can see by comparing the Figs. \ref{Fig:Entropia}(a) and \ref{Fig:CalorEspecifico}(a). Indeed, in the low-temperature, $C_{e}\simeq S$, with $S$ given by Eq. (\ref{Eq:entropia2}).
Therefore, we recover the results from the literature, namely, the heat capacity is proportional to both the density of states and the temperature \cite{kittel2005introduction}. 
However, we can see that additional oscillations appear when a subband is depopulated. Figures  \ref{Fig:CalorEspecifico}(b) and \ref{Fig:CalorEspecifico}(c) show this behavior more clearly. Oh et al. also observed this oscillatory behavior when studying the heat capacity of quantum wires and spherical dots \cite{JKPS.1995.28.132}.

In a subband, the excited states in a thickness $k_{B}T$ below the chemical potential and the upper levels above the chemical potential contribute separately to the heat capacity. When the chemical potential approaches the bottom of a subband, there is a reduction in the number of states below the chemical potential, which implies a decrease in heat capacity.
In this process, the number of available states above the chemical potential does not decrease.
Consequently, as soon as the contribution of these states prevails, the heat capacity increases again.
However, as the magnetic field increases and the chemical potential moves away from the bottom of a subband, the number of available states above the chemical potential also decreases. This, again, leads to a decrease in heat capacity.
As the temperature increases, Fig. \ref{Fig:CalorEspecifico}(d) shows another very interesting behavior in the heat capacity of quantum dots in the range of strong magnetic fields, namely,  the heat capacity presents a bump followed by a decrease until an apparent saturation value.
The saturation value corresponds to $Nk_{B}$, where $N$ is the number of electrons in the sample. At low temperatures, the value $Nk_{B}$ is reached in very strong magnetic fields. Knowing that heat capacity is an increasing function of temperature, we can also say that, in strong magnetic fields, heat capacity at low (high) temperatures increases (decreases) with increasing $B$. This result also makes it very clear how the magnetic field changes the dependence of heat capacity on temperature.
These results can also be seen in Ref. \cite{Entropy.2018.20.557}, 
the authors use the partition function to explore the thermodynamic properties of quantum dots as a temperature function for different magnetic field values.

\section{Conclusions}
\label{sec:Conc}

In this study, we have developed a model from which we extract the density of states of a 2DEG confined in a quantum dot. To test the model's validity, we compared our approach with known results from the literature and verified great accuracy. 

%In this study, we have developed a theoretical model to analyze the thermodynamic properties of a quantum dot containing confined electrons under the influence of an external magnetic field. By solving the Schrödinger equation with minimal coupling to a confinement potential and the magnetic field, we explored the effects of varying temperature and magnetic field on the system's energy, magnetization, entropy, and heat capacity.

One of the key observations of this study is the richness of phenomena observed in the thermodynamic properties of the quantum dot. We found that the presence of a magnetic field leads to distinct oscillations in the system's chemical potential, magnetization, entropia, and heat capacity, resulting from the depopulation of energy subbands as the magnetic field is increased. These oscillations reveal important information about the system's energy structure and are crucial for understanding the physics of nanostructured systems.

Furthermore, we observed that temperature plays a significant role in the thermodynamic properties of the system. Temperature variation influences the distribution of electrons in different energy subbands, directly affecting the entropy and heat capacity of the system. Notably, we observed phenomena such as the magnetocaloric effect, where the applied magnetic field modulates the system's temperature. In addition, two interesting aspects are observed in the heat capacity. First is the appearance of additional oscillations as the temperature increases. The second is the behavior of the heat capacity of a quantum dot when subjected to strong magnetic fields. These aspects suggest complex behaviors in the distribution of energy of confined electrons.

In summary, this study provides valuable insights into the effects of temperature and magnetic field on the thermodynamic properties of quantum dots. Additionally, the observations and results obtained here may be useful for developing new technologies in quantum devices and for the fundamental understanding of the physics of nanostructured systems.

\section*{Acknowledgments}
This work was partially supported by the Brazilian agencies CAPES, CNPq, and FAPEMA. E. O. Silva acknowledges CNPq Grant 306308/2022-3, FAPEMA Grants UNIVERSAL-06395/22 and APP-12256/22. This study was financed in part by the Coordena\c{c}\~{a}o de Aperfei\c{c}oamento de Pessoal de N\'{\i}vel Superior - Brasil (CAPES) - Finance Code 001.

\bibliographystyle{apsrev4-2}
%\bibliography{References}

\end{document}